\documentclass[floatfix,10pt,onebcolumn,showpacs,amsmath,amssymb]{revtex4}

\usepackage{epsf}
\usepackage{graphicx}  
\usepackage{dcolumn}   
\usepackage{bm}        

\newcommand{\be}{\begin{equation}}
\newcommand{\en}{\end{equation}}
 \newcommand{\bea}{\begin{eqnarray}}
 \newcommand{\ena}{\end{eqnarray}}
  \newcommand{\sch}{Schwarzschild}

\begin{document}

\title{Dynamical spacetimes in conformal gravity}
\author{Hongsheng Zhang $^{1,2,3~}$\footnote{Electronic address: hongsheng@shnu.edu.cn}, Yi Zhang $^4$, Xin-Zhou Li $^2$ }
\affiliation{$^1$ School of Physics and Technology, University of Jinan, West Road of Nan Xinzhuang 336,
Jinan, Shandong 250022, China\\
 $^2$ Center for Astrophysics, Shanghai Normal University, 100 Guilin Road,
Shanghai 200234, China\\
$^3$ State Key Laboratory of Theoretical Physics, Institute of Theoretical Physics, Chinese Academy of Sciences, Beijing, 100190, China\\
$^4$ College of Science, Chongqing University of Posts and Telecommunications, Chongqing 400065, China}

\date{ \today}

\begin{abstract}
   The conformal gravity remarkably boosts our prehension of gravity theories.  We find a series of dynamical solutions in the $W^2$-conformal gravity, including generalized \sch -Friedmann-Robertson-Walker (GSFRW), charged generalized \sch -Friedmann-Robertson-Walker (CGSFRW), especially rotating Friedmann-Robertson-Walker (RFRW),
   charged rotating Friedmann-Robertson-Walker (CRFRW), and a dynamical cylindrically symmetric solutions. The RFRW,
   CRFRW and the dynamical cylindrically symmetric solutions are never found in the Einstein gravity and modified gravities. The GSFRW and CGSFRW solutions take
   different forms from the corresponding solutions in the Einstein gravity.

\end{abstract}


\preprint{arXiv: }
 \maketitle

\section{Introduction}
  In 1918, Weyl proposed the purely infinitesimal geometry to unify gravity and electromagnetism \cite{wey}, the only two known interactions that time.
   Though Weyl's original suggestion to unify gravity and electromagnetic fields fails,
   the essential idea of the Weyl theory, i.e., the conformal symmetry, becomes more and more important in modern physics, for a modern review see Scholz \cite{olz}.
   The massless particles are conformally invariant. In sufficient high energy region, classically all the fields
   in the standard model are conformally invariant. Based on these facts, the conformal field theory is extensively applied
   in string theory, statistical physics, and condensed matter physics. However, gravity theory makes a significant exception. The Einstein gravity is not conformally invariant no matter how high the energy is. Imposing the
   condition of conformal invariance, one finds an interesting pure geometric gravity,
   \be
   S_W=\int d^4x \sqrt{-g}~ W_{abcd}W^{abcd},
   \label{geoa}
   \en
   in which $W_{abcd}$ denotes the Weyl tensor and $\alpha$ is a constant.

    The $W^2$-conformal gravity has been investigated in many
   aspects, for a review see Mannheim \cite{man1}. For spherical case, the solution of conformal gravity is helpful to interpret the
   surplus rotation velocity of the celestial objects far from the center of galaxies \cite{man2}, which is assumed to be the effects of dark
   matters in general relativity. The conformal gravity is subtle for cosmology. For an exact Friedmann-Robertson-Walker (FRW) universe, the Weyl
   tensor  always vanishes. An arbitrary scale factor is a solution of a void universe. The merit is that we do not need any exotic matter to
   explain the acceleration of the universe. The shortcoming is that an arbitrary scale factor makes the theory lose the ability of
   prediction. Fortunately the perturbations of matters in the universe will yield a non-zero Weyl tensor, and thus can present
   non-trivial constraints on the dynamical behaviours of the universe \cite{man3}. A data fitting of the cosmology in
   conformal gravity with perturbations has been done by Yang et al \cite{yang}.

    The boundary term of conformal gravity in the scenario of holography is presented by Grumiller et al \cite{gru}.
   The mass of any spacetime which is conformal to the Minkowski spacetime, for example an FRW universe, should be zero, which makes the concept of
   gravity mass in conformal gravity does not coincide with intuition \cite{des}. The conformal gravity is power-counting renormalizable. Thus, it is treated as a proper candidate for   quantum gravity theory. Recently 't Hooft presents an interesting argument how the requirement of renormalization leads to the conformal
   gravity from the Einstein gravity \cite{hoo}. All the vacuum solutions of the Einstein gravity are the vacuum solutions of the conformal
   gravity. The conformal gravity has more vacuum solutions, which do not satisfy the Einstein equation. We call such solutions ``non-trivial
    solutions" in conformal gravity. Maldacena shows that we can pick out
   the solutions of the Einstein equations from the solutions of conformal gravity by imposing a special boundary condition \cite{mal}.
   Liu et al find several interesting non-trivial solutions in conformal gravity \cite{liu}. Meng and Zhao find the C-metric solution in conformal gravity \cite{zhao}. The cylindrical solution in conformal gravity
   is studied by Said et al \cite{said}. We shall investigate the dynamical cylindrical solutions in this paper.

   A general conformally invariant gravity can be constructed by introducing an extra dilaton field. Narlikar et al showed that a wide class of singularities
   can be removed by choosing a ``physical frame" in the conformal gravity with a dilaton field \cite{nar} in cosmology. In black hole physics, the conformal invariance is also helpful to remove the spacetime singularities, see recent works by Bambi et al \cite{bambi1}, \cite{bambi2}, and Modesto et al \cite{modesto}.

    In this paper, we work in the frame of geometric conformal gravity (\ref{geoa}). This paper is organized as follows.  In section 2 we study the GSFRW solution and the CGSFRW solution. Similar solutions have been studied in
    the Einstein gravity. Here they appear in different form with vanishing energy-stress. In section 3 we obtain the RFRW and CRFRW solutions, which
    never found in the Einstein gravity and modified gravities. In section 4, we present  special dynamical cylindrical solutions. They are  non-trivial solutions in conformal gravity. In section 5, we conclude this paper.

    \section{GSFRW and CGSFRW spacetimes in conformal gravity}
    To find an exact solution of a mass point in a dynamical universe is an old problem. In 1933, McVittie found a solution which can describe the
    gravitational field of a mass point in an FRW universe \cite{mcv}. Recently, Gao et al found the higher dimensional McVittie solution and charged FRW solution
    in the Einstein gravity \cite{gao}. Here, we try to find the corresponding solutions in conformal gravity.

    \subsection{GSFRW}
    We will show that the McVittie metric and Gao's charged FRW metric are not solutions of conformal gravity. Hence it is a non-trivial work to
    find the GSFRW and CGSFRW in conformal gravity.

    The McVitte solution reads,
    \be
   ds^2= -\frac{\left(\sqrt{a(t)}-\frac{M}{2r \sqrt{a(t)}}\right)^2}{\left(\frac{M}{2r \sqrt{a(t)}}+\sqrt{a(t)}\right)^2} dt^2+\left(\frac{M}{2r
   \sqrt{a(t)}}+\sqrt{a(t)}\right)^4 \left(dr^2+r^2 d\theta^2+r^2 \text{sin}^2\theta d\phi^2\right),
    \label{mcvt}
        \en
  where $M$ is the \sch ~mass parameter, $a(t)$ denotes the scale factor, and $t,~r,~\theta,~\phi$ are the isotropic spherical coordinates. The action of the
   conformal gravity with matter reads,
   \be
   S=\int d^4x \sqrt{-g}~ \left(-\frac{\alpha}{16\pi} W_{abcd}W^{abcd}+{\cal L}_m\right),
   \label{act}
   \en
   where ${\cal L}_m$ denotes the Lagrangian of the matter fields. The corresponding equation of field reads,
   \be
   -4\alpha C_{ab}=8\pi T_{ab},
   \label{bach}
   \en
   where $T_{ab}$ is the stress-energy corresponding the ${\cal L}_m$, and $C_{ab}$ is the Bach tensor,
   \be
   C_{ab}=(\nabla^c\nabla^d+\frac{1}{2}R^{cd})W_{acbd}.
   \en
      Substituting the McVittie metric (\ref{mcvt}) into the field equation (\ref{bach}) and assuming the stress energy is in perfect fluid form, we derive
   \be
   \rho=-T_0^0=0,
   \en
   \be
  8\pi p_r=8\pi T_1^1=\frac{16 \alpha M r^3 a \left(-\dot{a}^2+a \ddot{a}\right)}{(M-r a) (M+r a)^5},
   \en
   \be
  8\pi p_\theta=8\pi p_\phi=8\pi T_2^2=8\pi T_3^3=-\frac{8 \alpha M r^3 a \left(-\dot{a}^2+a \ddot{a}\right)}{(M-r a) (M+r a)^5}.
   \en
   It is clear that $p_r\neq p_\theta$, and thus the universe is anisotropic. This form of pressure violates
   the fundamental motivation of McVittie to find his solution. Furthermore, a universe with zero density but
   finite pressure is very difficult to understand. This leads to the conclusion that the McVittie solution is not
   the SFRW solution in conformal gravity. From the other point of view, the stress energy tensor of the matters in conformal gravity must be
   traceless. It is clear that the McVittie metric does not satisfy this requirement. Thus, it is not a solution of conformal gravity.

   Now we try to find the GSFRW solution in conformal gravity. Both for the \sch~and FRW solutions the stress-energy
   vanish in conformal gravity. This condition determines the asymptotic behaviour of the metric. Generally,
   the metric ansatz can be written as,
   \be
   ds^2=-f(r)F(t,r)dt^2+B^2(t,r)\left(f^{-1}(r)dr^2+r^2 d\theta^2+r^2 \text{sin}^2\theta d\phi^2\right).
   \label{cross}
   \en
  Here we write the metric in a \sch-like form rather than the isotropic form. The convenience of such
  coordinates will be clear later. The corresponding field equation of the metric (\ref{cross})
  is very involved. We cast it into the Appendix A. It is difficult to find a general solution for $f$, $F$ and
  $B$.  Here we study an explicit and physically more important case for $F$ and $B$,
  \be
  F=F(r),~~~B=B(t).
  \label{orth}
  \en
  Then we obtain the field equations as follows,
  \be
  -8\pi \rho=-\frac{\alpha}{24 r^4 B^4 F^4}\left(49 r^4 f^2 F'^4-2 r^3 f F F'^2 \left(47 r f' F'+f \left(22 F'+58 r F''\right)\right)+r^2 F^2 \left(25 r^2 f'^2 F'^2 \right.\right.
  \nonumber
  \en
  \be
  \left.+4 r f F' \left(13 r F' f''+f' \left(31 F'+41 r F''\right)\right)+4 f^2 \left(-5 F'^2+9 r^2 F''^2+2 r F' \left(13 F''+6 r F'''\right)\right)\right)
  \nonumber
  \en
  \be
  +4 F^4 \left(-4+4 f^2+4 r^2 f'^2+r^4 f''^2-2 r^3 f' \left(2 f''+r f'''\right)-4 r f \left(2 f'+r \left(-f''+3 r f'''+r^2 f''''\right)\right)\right)
  \nonumber
  \en
  \be
  \left.-4 r^3 F^3 \left(5 f'^2 \left(2 F'+r F''\right)+16 f f' \left(3 F''+r F'''\right)+2 f \left(7 r f'' F''+F' \left(14 f''+3 r f'''\right)+8 f F'''+2 r f F''''\right)\right)\right)
  \en
  \be
  8\pi p_r=\frac{\alpha}{24 r^4 B^4 F^4}\left(16 F^4-16 f^2 F^4+32 r f F^4 f'-16 r^2 F^4 f'^2+32 r f^2 F^3 F'-80 r^2 f F^3 f' F'+40 r^3 F^3 f'^2 F'-4 r^2 f^2 F^2 F'^2 \right.
    \nonumber
  \en
    \be
  +52 r^3 f F^2 f' F'^2-25 r^4 F^2 f'^2 F'^2-20 r^3 f^2 F F'^3-2 r^4 f F f' F'^3+7 r^4 f^2 F'^4-16 r^2 f F^4 f''+16 r^3 F^4 f' f''-16 r^3 f F^3 F' f''
  \nonumber
  \en
  \be
  +16 r^4 f F^2 F'^2 f''-4 r^4 F^4 f''^2-32 r^2 f^2 F^3 F''-16 r^3 f F^3 f' F''+20 r^4 F^3 f'^2 F''+48 r^3 f^2 F^2 F' F''-8 r^4 f F^2 f' F' F''
  \nonumber
  \en
  \be
  -12 r^4 f^2 F F'^2 F''-8 r^4 f F^3 f'' F''-4 r^4 f^2 F^2 F''^2-16 r^3 f F^4 f'''+8 r^4 F^4 f' f'''+8 r^4 f F^3 F' f'''-16 r^3 f^2 F^3 F'''
  \nonumber
  \en
  \be
  \left.+8 r^4 f F^3 f' F'''+8 r^4 f^2 F^2 F' F'''\right)
  \en
  \be
 8\pi p_\theta=8\pi p_\phi=\frac{\alpha}{24 r^4 B^4 F^4}\left(-16 F^4+16 f^2 F^4-32 r f F^4 f'+16 r^2 F^4 f'^2-16 r f^2 F^3 F'+40 r^2 f F^3 f' F'-40 r^3 F^3 f'^2 F'
  \right.
  \nonumber
  \en
  \be
  -8 r^2 f^2 F^2 F'^2+36 r^3 f F^2 f' F'^2+25 r^4 F^2 f'^2 F'^2-12 r^3 f^2 F F'^3-46 r^4 f F f' F'^3+21 r^4 f^2 F'^4+16 r^2 f F^4 f''-16 r^3 F^4 f' f''
  \nonumber
  \en
  \be
  -48 r^3 f F^3 F' f''+18 r^4 f F^2 F'^2 f''+4 r^4 F^4 f''^2+16 r^2 f^2 F^3 F''-88 r^3 f F^3 f' F''-20 r^4 F^3 f'^2 F''+28 r^3 f^2 F^2 F' F''
  \nonumber
  \en
  \be
  +86 r^4 f F^2 f' F' F''-52 r^4 f^2 F F'^2 F''-24 r^4 f F^3 f'' F''+20 r^4 f^2 F^2 F''^2-16 r^3 f F^4 f'''-8 r^4 F^4 f' f'''-16 r^4 f F^3 F' f'''
  \nonumber
  \en
  \be
  \left.
  -24 r^3 f^2 F^3 F'''-36 r^4 f F^3 f' F'''+20 r^4 f^2 F^2 F' F'''-8 r^4 f F^4 f''''-8 r^4 f^2 F^3 F''''\right)
  \en
  To obtain a reasonable solution, we further make two assumptions. The first one is the density is independent of $r$,
  \be
  \rho=\rho(t).
  \label{erho}
  \en
  This requirement is inspired by the McVittie solution in the Einstein gravity. The second is the universe is isotropic. We should be cautious
  when using this condition. When we say that the pressure is isotropic we usually require that the pressure is measured
  in the isotropic coordinates. But we are working in  the \sch-like coordinates. Now we prove that the pressures in different directions are not changed by coordinates
  transformation in the following form,
  \be
  \bar{t}=t,~~~\bar{r}=\bar{r}(r),~~~\bar{\theta}=\theta,~~~\bar{\phi}=\phi.
  \label{tra}
  \en
  In the bar coordinates, the pressure in the $\bar{r}$-direction reads,
  \be
  \bar{T}^1_1=T^{\mu}_{\nu}\frac{\partial \bar{x}^1}{\partial x^{\mu}}\frac{\partial {x}^\nu}{\partial \bar{x}^{1}}=
  T^{1}_{1}\frac{\partial \bar{r}}{\partial r}\frac{\partial {r}}{\partial \bar{r}}=T^1_1.
    \en
  Similarly, we can demonstrate that $p_\theta$ and $p_\phi$ are coordinate-independent under the transformation of
  (\ref{tra}). We can obtain the isotropic coordinates through coordinates transformation as a special case of (\ref{tra}).
    Thus we prove that the isotropic condition is the same in  \sch-like coordinates as in the isotropic ones,
    \be
    p_r=p_\theta=p_\phi.
    \label{eiso}
    \en
    Since the Bach tensor $C_{ab}$ is traceless, the stress energy $T_{ab}$ in (\ref{bach}) is traceless, which requires
    \be
     -\rho+p_r+p_\theta+p_\phi=0.
     \label{TR}
     \en
  Solving the equations (\ref{erho}) and (\ref{eiso}), we obtain,
  \be
  F={\rm constant},
  \en
  and
  \be
  f=\frac{h}{r}+i+jr+kr^2,
  \label{fgg}
  \en
  where $h,~i,~j,~k$ are integration constants, and
  \be
  i^2-3hj=1.
  \en
   $B(t)$ is an arbitrary function of $t$. Without loss of generality we set $F=1$. In this solution the stress energy vanishes,
   \be
   \rho=0,~p_r=p_\theta=p_\phi=0,
   \en
   which is the simplest case satisfying the condition (\ref{TR}).
   The stress energy satisfies the required asymptotic behaviours of the GSFRW solution.  An empty universe is not completely nonsense. As an important toy model, the Milne universe has been discussed
  for many years \cite{mil}. Recently, Nielsen et al show that the dynamics of the universe may be exactly like
     an empty universe \cite{nie}. Furthermore, the perturbation of matters in the universe will yield a non-zero stress-energy in conformal
  gravity.

  On one hand, from the above discussions the conformal invariance imposes so harsh constraint on the matter distributions in the universe that
  it seems only a vanishing stress energy is permitted for isotropic distributions. On the other hand, it offers extra freedoms in the dynamics of the universe.
  In an empty FRW universe, the scale factor can be an arbitrary function of the cosmic time. We see that the GSFRW metric inherits this property. In fact, we can solve the metric ansatz (\ref{cross}) through the other road by taking full advantages of the conformal invariance. Here we present a sketch of this method.  First, we set
  \be
  F(t,r)=P(r)B^2(t,r).
  \en
  And then the metric is equal to the following metric in conformal gravity,
  \be
  ds^2=-f(r)P(r)dt^2+\left(f^{-1}(r)dr^2+r^2 d\theta^2+r^2 \text{sin}^2\theta d\phi^2\right).
   \label{crossp}
   \en
  By using a similar procedure in \cite{bir}, one can prove $P(r)=1$. Then restoring the scale factor $B$, and making a time rescaling, one can also arrive at the previous GSFRW metric.

  Now we discuss some special cases of the above solution. When $f$ degenerates to the \sch~form, we have $j=k=0,~i=1$, and thus the solution becomes,
  \be
  ds^2=-fdt^2+B^2(f^{-1}dr^2+r^2d\theta^2+r^2\sin^2\theta d\phi^2).
  \en
  In the isotropic coordinates, it reads,
  \be
  ds^2=-\frac{\left(1-\frac{M}{2r }\right)^2}{\left(1+\frac{M}{2r }\right)^2} dt^2+B^2\left(1+\frac{M}{2r}\right)^4 \left(dr^2+r^2 d\theta^2+r^2 \text{sin}^2\theta d\phi^2\right).
  \en
   Comparing with (\ref{mcvt}), one finds that it is still different from the McVitte solution. As we have mentioned, the
   McVitte metric is not a solution of the conformal gravity. So it is impossible to recover the McVitte metric from our solution.

   Now we make a short discussion about the isotropic coordinates. Generally, the isotropic coordinate is derived by
   \be
   \rho_i=\int \frac{~dr}{\sqrt{f}}.
   \en
   It has no analytical result for a general $f$ in (\ref{fgg}), which implies that we cannot obtain the analytical
   result if we work in the isotropic coordinates from the beginning. That is the reason why we use \sch-like coordinates
   in (\ref{cross}). The next important case is the static limit. When $B=$constant, one recovers the solution found in \cite{bir}.

   For a fixed $r$, the GSFRW metric describes an expanding/contracting two-sphere, which can be treated as a three-dimensional
   FRW universe. However, for a variable  $r$, it does not become a four-dimensional FRW universe when $r\to \infty$ except the case
   $j=k=0$. So a  proper name for this solution is generalized  FRW solution.
   \subsection{CGSFRW}
   Because the charged FRW solution in [13] admits the McVitte solution as a special case, it is not a solution of conformal gravity either.
   We try to find the CGSFRW solution in conformal gravity. When we turn on the (sourced) $U(1)$ gauge field, the matter lagrangian in (\ref{act})
   becomes
   \be
   {\cal L}_m={\cal \bar{L}}-\frac{1}{4}{\cal F}_{ab}{\cal F}^{ab}-A_aJ^a,
   \en
  where ${\cal F}=dA$ denotes the $U(1)$ gauge field, $J$ denotes the
  electromagnetic current, and ${\cal \bar{L}}$ represents the contribution of other matters.

   Similar to the previous discussions, we obtain a solution under the metric ansatz (\ref{cross}) and (\ref{orth}),
   \be
   F={\rm constant},
   \en
   \be
   f=\frac{h}{r}+i+jr+kr^2,
   \en
   where $h,~i,~j,~k$ are constants, and the stress energy corresponding to $\cal \bar{L}$ vanishes.
   The stress energy for this metric reads,
   \be
   8\pi \rho=-8\pi T_0^0=-\frac{2\alpha(1- i^2+3 h j)}{3 r^4 B^4},
   \en
   \be
   8\pi p_r=8\pi T_1^1=\frac{2\alpha(1- i^2+3 h j)}{3 r^4 B^4},
   \en
   \be
   8\pi p_\theta=8\pi p_\phi=8\pi T_2^2=8\pi T_3^3=-\frac{2\alpha(1- i^2+3 h j)}{3 r^4 B^4}.
   \en
   One can check that the stress energy is traceless,
   \be
   T^\mu_{\mu}=0.
   \en
   The potential of the U(1) field is,
   \be
   A=\frac{q(t)}{r}dt-v\cos\theta d\phi.
   \label{pot}
   \en
  Here $q(t),~v$ are the electric and the magnetic charge of the point mass at $r=0$, respectively.
  The stress energy of the U(1) gauge field reads,
  \be
  8\pi T^0_0(U(1))=-\frac{  \left(v^2+B^2 q^2\right)}{ r^4 B^4},~T^1_1(U(1))=-T^2_2(U(1))=-T^3_3(U(1))=T^0_0(U(1)).
  \en

  Then the field equation requires
  \be
  Bq=K,
  \label{bq}
  \en
  and
  \be
  K^2=\frac{2\alpha(i^2-1-3hj)}{3 }-v^2.
  \en
  The total charge in a 2-surface with radius $r$ is derived by the integration of the Hodge dual of the field strength,
  \be
  Q=\frac{1}{4\pi}\int *{\cal F}=K,
  \en
  which is independent on $r$. This is the property of a point charge. One sees that the total charge is
  time independent, which is different from $q(t)$. This point also can be seen from the electric current reads,
  \be
  J=\frac{1}{r^2fB }(q\dot{B}+B\dot{q})dr=0
  \en
  Thus the charge conservation law requires the central charge to be a constant.
  The potential (\ref{pot}) becomes
   \be
   A=\frac{K}{Br}dt-v\cos\theta d\phi.
   \label{pot1}
   \en
   At the first sight, this is a very natural, even obvious result. The electric potential decreases to $1/B$ when the universe
   expand by a factor $B$ for a constant charge. In fact, it is a highly non-trivial result, since the stress energy must
   increase or decrease synchronously to satisfy the field equation. In fact this point cannot be accomplished in the Einstein gravity \cite{gao},
   in which a variable charge, and thus a non-zero current, are required.

   At the limit $B$=constant, such
  that $q$=constant, and the magnetic charge $v$ vanishes, the previous result in \cite{bir} recovers.

  The metric cannot come back to the Reissner-Nordstrom metric with any special choices of parameters.
  It really describes an asymptotic FRW  spacetime with a point charge. So we name it CGSFRW metric in conformal gravity.
  \section{RFRW and CRFRW spacetimes in conformal gravity}
 To describe a rotating black hole in an FRW universe is a long-waited achievement. Unfortunately, the Kerr-FRW metric has not been found in
 the Einstein gravity up to now. And to our knowledge, it has never been found in any modified gravity theory. By using the
  special structure of the conformal gravity, we successfully find the RFRW and CRFRW metrics.

  To derive the Kerr solution in the Einstein gravity is a very complicated work. The field equation becomes extremely
  complicated when one considers a rotating solution which is asymptotic to the FRW metric. In this section we just
  present our results. One can verify that the following metric solves the field equation of the conformal gravity with a $U(1)$ gauge field,
  \be
  ds^2=B(t)^2 \left(r^2+a^2 \cos^2\theta\right) \left(\frac{1}{a^2+r^2-2 r \left(M+c r^2\right)} dr^2+\frac{1}{1+b \cos\theta} d\theta^2  \right)
  \nonumber
  \en
  \be
  +\frac{1+b \cos\theta}{r^2+a^2 \cos^2\theta} \sin^2\theta \left(adt-\left(a^2+r^2\right) B(t) d\varphi  \right)^2-\frac{a^2+r^2-2 r \left(M+c r^2\right)}{r^2+a^2 \cos^2\theta^2} \left(dt-a B(t) \sin^2\theta d\varphi\right)^2.
  \label{kfrw}
  \en
  Here, $a$ denotes the angular momentum per mass, $M$ is a mass parameter, $B(t)$ is a function of time, which reduces to the scale factor of the \sch-FRW universe in \sch-like coordinate system when $a=0,~b=0,~c=0$, and $b$ and $c$ are two extra parameters which have no correspondence in the Einstein gravity. It is easy
  to check that it comes back to the Kerr metric when $b=c=0$ and $B$ is a constant. The stress energy of the solution (\ref{kfrw}) reads,
    \be
  -8\pi \rho=8\pi T^0_0=\frac{8 \alpha\left(b^2+4 c M\right) \left(3 a^2+2 r^2-a^2 \cos2\theta\right)}{B^4\left(a^2+2 r^2+a^2 \cos2\theta\right)^3}
  \en
   and
   \be
  8\pi p_1=8\pi T_1^1=\frac{2\alpha \left(b^2+4 c M\right)}{B^4 \left(r^2+a^2 \cos^2\theta\right)^2},
   \en
   \be
    p_2=- p_1=T^2_2, p_3=T_3^3=-T^0_0.
   \en
   One sees that the components of the stress energy satisfy,
   \be
   T_3^3+T^0_0=0;~T_1^1+T_2^2=0.
   \en
   Thus it satisfies the requirement of the stress energy in conformal gravity.
   The potential of the $U(1)$ gauge field reads,
   \be
   A=\frac{Kr}{\rho B}(dt-a\sin^2\theta B d\varphi)+\frac{v\cos\theta}{\rho B}(adt-B(r^2+a^2)d\varphi),
   \en
  whose energy stress reads,
  \be
 8\pi T^0_0=-\frac{4   \left(K^2+v^2\right) \left(3 a^2+2 r^2-a^2 \cos2\theta\right)}{B^4 \left(a^2+2 r^2+a^2 \cos2\theta\right)^3}
  \en
  \be
  8\pi T^1_1=-\frac{  \left(K^2+v^2\right)}{ B^4 \left(r^2+a^2 \cos^2\theta\right)^2},
  \en
  \be
  T_2^2=-T^1_1,~~~T^3_3=-T^0_0.
  \en
 The field equation requires,
 \be
  (K^2+v^2)=-2\alpha (b^2+4cM).
 \en

 Similar to  the demonstrations in the above context, one can show that $K$ is the total electric charge and $v$ is the
 total magnetic charge in the universe. Also, this point can be confirmed by the fact that the electromagnetic currents vanish.
 Thus generally, it describes a charged rotating asymptotic FRW universe in conformal gravity. When $b^2+4cM=0$, we obtain a neutral universe. And furthermore, we obtain a Kerr-FRW universe if we set $b=0,~c=0$. The Kerr-FRW universe
 is never obtained in the Einstein gravity, nor in any modified gravity theory.

  \section{dynamical cylindrically symmetric solution}
  The general line element of a cylindrically symmetric solution can be written as,
  \be
  ds^2=-H_0dt^2+H_1dz^2+H_2dr^2+H_3d\phi^2,
  \en
  where $H_0,~H_1,~H_2,~H_3$ are functions of $r$ and $t$. The cylindrically symmetric solutions are
  widely investigated in the Einstein gravity. But, the dynamical solutions are seldom studied. We shall
  study some cylindrically dynamical solutions in conformal gravity.

  A class of static  cylindrically symmetric solution has been found in \cite{said},
  \be
   ds^2=-f(r)dt^2+f^{-1}(r)dr^2+r^2 dz^2+r^2 \beta d\phi^2,
   \label{cyli}
   \en
  in which the constant $\beta$ denotes the possible conical singularity, and
  \be
  f=\frac{h}{r}+i+jr+kr^2.
  \label{cyli2}
  \en
  This looks very similar to the spherical solution in the conformal gravity. The different points are discussed
  as follows. The first one is that the subspace ($z,~\phi$) is flat, while the subspace ($\theta,~\phi$) is curved in
  the  spherical case. The second one is that the stress energy vanishes, while a stress energy of a U(1) gauge field
  appears if one does not impose any constraints among the constants $h,~i,~j,$ and $k$ in the case of the spherical solution.
  Actually,  (\ref{cyli}) also can be treated as a topological solution corresponding to the spherical solution when $\beta=1$, like
  the topological \sch~ solution in the Einstein gravity.

  Then we discuss the dynamical case.
  Mimicking the spherical case, we derive the following dynamical solution in conformal gravity
 \be
   ds^2=-f(r)dt^2+B(t)^2(f^{-1}(r)dr^2+r^2 dz^2+r^2 \beta d\phi^2).
   \label{cds}
   \en
 The stress energy in this solution is zero. So it is a vacuum solution which describes an expanding/contracting cylindrically symmetric
 space.

 When $i=j=0$, (\ref{cyli2}) becomes
 the Melvin solution \cite{mel}.  We find the other type of dynamical solution of with dynamical subspace of ($z,~\phi$) belonging
 to the Melvin class,

 \be
   ds^2=-f(r)dt^2+f^{-1}(r)dr^2+B^2(r^2 dz^2+r^2 \beta d\phi^2),
   \label{cds2}
   \en
 in which $i=j=0,$  $hk=0$ and $B=t^{3/4}$. This solution describes a cylindrically symmetric with an expanding subspace ($z,~\phi$).
 \section{conclusions and discussions}

 The conformal gravity takes a special status in modified gravity theories. Classically, the standard model of particle physics is conformally
 invariant at high energy region. If the quantum effects are considered, a conformal anomaly appears \cite{modesto}. In classical level, if the putative unified theory including gravity and gauge interactions
 exists, the gravity may also be conformally invariant. However, classically the Einstein gravity is not
  conformally invariant. A pure geometric Lagrangian of  conformally invariant gravity in four dimensional
 spacetime is the contraction of the Weyl tensor. We study some dynamical solutions in such a conformal gravity.

 All the vacuum solutions of the Einstein gravity are the solutions of conformal gravity. A sourced solution in the Einstein gravity
 may not solve the conformal gravity. For example, a sourced FRW universe is not a solution in conformal gravity. The FRW metric, as
 a solution, is a vacuum solution in conformal gravity. We show that the \sch-FRW metric (the McVittie metric) is not a solution of conformal gravity.
 In this article we find the GSFRW solution in conformal gravity. And then, we extend it to the charged case.

 To describe a rotating black hole inhabited in an FRW universe is an extremely difficult problem in gravity theory. Before this research, there is
 no exact solution to describe a rotating black hole in an FRW universe in the Einstein gravity and in modified gravity theories. We find a class
 of rotating solution in an FRW universe in conformal gravity. Both the neutral and charged cases are considered. Finally, we investigate
 the dynamical cylindrically symmetric solutions. There are several other methods to obtain the exact solutions besides directly solving the field equation.
 A noticeable recent approach is the thermodynamic method \cite{self1}. Also, there are several solutions with different symmetries in the Einstein gravity, for example the plane symmetric solutions \cite{self2}. How to extend these method and solutions to the case of the conformal gravity is the future topic.

 {\bf Acknowledgments.}
   The authors thank the anonymous referee for her/his valuable comments. This work is supported by the Program for Professor of Special Appointment (Eastern Scholar) at Shanghai Institutions of Higher Learning, National Education Foundation of China under grant No. 200931271104, Shanghai Key Laboratory of Particle Physics and Cosmology under grant No.11DZ2230700, and National Natural Science Foundation of China under Grant Nos. 11075106, 11275128 and 11105004.

  \appendix{\bf{Appendix A}}

  The field equations under the ansatz (\ref{cross}) read,
  \be
  -8\pi \rho=\frac{\alpha}{24 r^4 B^6 F^4}\left(135 r^4 f^2 F^4 B'^4-10 r^3 B f F^3 B'^2 \left(15 r F f' B'+2 f \left(r B' F'+F \left(11 B'+13 r B''\right)\right)\right)\right.
  \nonumber
  \en
  \be
  +5 r^2 B^2 F^2 \left(3 r^2 F^2 f'^2 B'^2+2 r f F B' \left(r f' B' F'+F \left(6 r f'' B'+f' \left(26 B'+22 r B''\right)\right)\right)\right.
  \nonumber
  \en
  \be
   \left.\left.\left.+2 f^2 \left(-3 r^2 B'^2 F'^2+2 r F B' \left(2 r F' B''+B' \left(5 F'+r F''\right)\right)\right.\right.+2 F^2 \left(B'^2+3 r^2 B''^2+2 r B' \left(7 B''+2 r B'''\right)\right)\right)\right)
  \nonumber
  \en
  \be
  -2 r^2 B^3 F \left(18 r^2 f^2 B' F'^3-r f F F' \left(23 r f' B' F'+2 f \left(7 r F' B''+B' \left(19 F'+14 r F''\right)\right)\right)\right.
  \nonumber
  \en
  \be
  +F^2 \left(-5 r^2 f'^2 B' F'+2 r f \left(2 r f'' B' F'+f' \left(7 r F' B''+B' \left(34 F'+7 r F''\right)\right)\right)+4 f^2 \left(r \left(3 r B'' F''+F' \left(11 B''+2 r B'''\right)\right)\right.\right.
  \nonumber
  \en
  \be
  +B' \left.\left.\left(5 F'+r \left(11 F''+2 r F'''\right)\right)\right)\right)+2 r F^3 \left(5 f'^2 \left(2 B'+r B''\right)+16 f f' \left(3 B''+r B'''\right)\right.
  \nonumber
  \en
  \be
  +2 f \left.\left.\left(3 r f''' B'+7 f'' \left(2 B'+r B''\right)+8 f B'''+2 r f B''''\right)\right)\right)
  \nonumber
  \en
  \be
  +B^4 \left(-4 F^4 \left(-4+4 f^2+4 r^2 f'^2+r^4 f''^2-2 r^3 f' \left(2 f''+r f'''\right)-4 r f \left(2 f'+r \left(-f''+3 r f'''+r^2 f''''\right)\right)\right)\right.
  \nonumber
  \en
  \be
  -49 r^4 f^2 F'^4+2 r^3 f F F'^2 \left(47 r f' F'+f \left(22 F'+58 r F''\right)\right)-r^2 F^2 \left(25 r^2 f'^2 F'^2+4 r f F' \left(13 r f'' F'+f' \left(31 F'+41 r F''\right)\right)\right.
  \nonumber
  \en
  \be
  +4 f^2 \left.\left(-5 F'^2+9 r^2 F''^2+2 r F' \left(13 F''+6 r F'''\right)\right)\right)+4 r^3 F^3 \left(5 f'^2 \left(2 F'+r F''\right)+16 f f' \left(3 F''+r F'''\right)\right.
  \nonumber
  \en
  \be
  +2 f\left.\left.\left. \left(3 r f''' F'+7 f'' \left(2 F'+r F''\right)+8 f F'''+2 r f F''''\right)\right)\right)\right),
  \en

  \be
   8\pi p_r=\frac{\alpha}{24 r^4 B^6 f F^5}\left(16 B^4 f F^5-16 B^4 f^3 F^5+32 r B^4 f^2 F^5 f'-16 r^2 B^4 f F^5 f'^2-16 r^2 B^4 f^2 F^5 f''+16 r^3 B^4 f F^5 f' f''\right.
    \nonumber
  \en
  \be
        -4 r^4 B^4 f F^5 f''^2-16 r^3 B^4 f^2 F^5 f'''+8 r^4 B^4 f F^5 f' f'''-32 r B^3 f^3 F^5 B'+80 r^2 B^3 f^2 F^5 f' B'-40 r^3 B^3 f F^5 f'^2 B'+16 r^3 B^3 f^2 F^5 f'' B'
        \nonumber
  \en
  \be
        -8 r^4 B^3 f^2 F^5 f''' B'-68 r^2 B^2 f^3 F^5 B'^2+20 r^3 B^2 f^2 F^5 f' B'^2+15 r^4 B^2 f F^5 f'^2 B'^2+20 r^3 B f^3 F^5 B'^3-30 r^4 B f^2 F^5 f' B'^3
        \nonumber
  \en
  \be
        +15 r^4 f^3 F^5 B'^4+32 r B^4 f^3 F^4 F'-80 r^2 B^4 f^2 F^4 f' F'+40 r^3 B^4 f F^4 f'^2 F'-16 r^3 B^4 f^2 F^4 f'' F'+8 r^4 B^4 f^2 F^4 f''' F'
        \nonumber
  \en
  \be
        +72 r^2 B^3 f^3 F^4 B' F'-72 r^3 B^3 f^2 F^4 f' B' F'+10 r^4 B^3 f F^4 f'^2 B' F'-16 r^4 B^3 f^2 F^4 f'' B' F'+36 r^3 B^2 f^3 F^4 B'^2 F'
        \nonumber
  \en
  \be
        +10 r^4 B^2 f^2 F^4 f' B'^2 F'-20 r^4 B f^3 F^4 B'^3 F'-4 r^2 B^4 f^3 F^3 F'^2+52 r^3 B^4 f^2 F^3 f' F'^2-25 r^4 B^4 f F^3 f'^2 F'^2
        \nonumber
  \en
  \be
       +16 r^4 B^4 f^2 F^3 f'' F'^2-36 r^3 B^3 f^3 F^3 B' F'^2+22 r^4 B^3 f^2 F^3 f' B' F'^2+2 r^4 B^2 f^3 F^3 B'^2 F'^2-20 r^3 B^4 f^3 F^2 F'^3
       \nonumber
  \en
  \be
       -2 r^4 B^4 f^2 F^2 f' F'^3-4 r^4 B^3 f^3 F^2 B' F'^3+7 r^4 B^4 f^3 F F'^4+32 r^2 B^3 f^3 F^5 B''+16 r^3 B^3 f^2 F^5 f' B''-20 r^4 B^3 f F^5 f'^2 B''
       \nonumber
  \en
  \be
       +8 r^4 B^3 f^2 F^5 f'' B''-48 r^3 B^2 f^3 F^5 B' B''+40 r^4 B^2 f^2 F^5 f' B' B''-20 r^4 B f^3 F^5 B'^2 B''-16 r^4 B^3 f^2 F^4 f' F' B''
       \nonumber
  \en
  \be
       +32 r^4 B^2 f^3 F^4 B' F' B''-12 r^4 B^3 f^3 F^3 F'^2 B''-4 r^4 B^2 f^3 F^5 B''^2-32 r^2 B^4 f^3 F^4 F''-16 r^3 B^4 f^2 F^4 f' F''
       \nonumber
  \en
  \be
       +20 r^4 B^4 f F^4 f'^2 F''-8 r^4 B^4 f^2 F^4 f'' F''-16 r^4 B^3 f^2 F^4 f' B' F''-4 r^4 B^2 f^3 F^4 B'^2 F''+48 r^3 B^4 f^3 F^3 F' F''
       \nonumber
  \en
  \be
       -8 r^4 B^4 f^2 F^3 f' F' F''+16 r^4 B^3 f^3 F^3 B' F' F''-12 r^4 B^4 f^3 F^2 F'^2 F''+8 r^4 B^3 f^3 F^4 B'' F''-4 r^4 B^4 f^3 F^3 F''^2
       \nonumber
  \en
  \be
       +16 r^3 B^3 f^3 F^5 B'''-8 r^4 B^3 f^2 F^5 f' B'''+8 r^4 B^2 f^3 F^5 B' B'''-8 r^4 B^3 f^3 F^4 F' B'''-16 r^3 B^4 f^3 F^4 F'''
       \nonumber
  \en
  \be
       +8 r^4 B^4 f^2 F^4 f' F'''-8 r^4 B^3 f^3 F^4 B' F'''+8 r^4 B^4 f^3 F^3 F' F'''-24 r^3 B^2 f F^4 B' \dot{B}^2+36 r^4 B^2 F^4 f' B' \dot{B}^2
       \nonumber
  \en
  \be
       -120 r^4 B f F^4 B'^2 \dot{B}^2+24 r^4 B^2 f F^3 B' F' \dot{B}^2+24 r^4 B^2 f F^4 B'' \dot{B}^2-8 r^3 B^3 f F^3 B' \dot{B} \dot{F}+12 r^4 B^3 F^3 f' B' \dot{B} \dot{F}
       \nonumber
  \en
  \be
       -24 r^4 B^2 f F^3 B'^2 \dot{B} \dot{F}-8 r^3 B^4 f F^2 F' \dot{B} \dot{F}+12 r^4 B^4 F^2 f' F' \dot{B} \dot{F}+32 r^4 B^3 f F^2 B' F' \dot{B} \dot{F}-8 r^4 B^4 f F F'^2 \dot{B} \dot{F}
       \nonumber
  \en
  \be
       +8 r^4 B^3 f F^3 B'' \dot{B} \dot{F}+8 r^4 B^4 f F^2 F'' \dot{B} \dot{F}+40 r^3 B^5 f F F' \dot{F}^2-60 r^4 B^5 F f' F' \dot{F}^2+40 r^4 B^4 f F B' F' \dot{F}^2
       \nonumber
  \en
  \be
       +56 r^4 B^5 f F'^2 \dot{F}^2-40 r^4 B^5 f F F'' \dot{F}^2+24 r^3 B^3 f F^4 \dot{B} \dot{B}'-36 r^4 B^3 F^4 f' \dot{B} \dot{B}'+168 r^4 B^2 f F^4 B' \dot{B} \dot{B}'
       \nonumber
  \en
  \be
       -24 r^4 B^3 f F^3 F' \dot{B} \dot{B}'+8 r^3 B^4 f F^3 \dot{F} \dot{B}'-12 r^4 B^4 F^3 f' \dot{F} \dot{B}'+24 r^4 B^3 f F^3 B' \dot{F} \dot{B}'
       \nonumber
  \en
  \be
       -40 r^4 B^4 f F^2 F' \dot{F} \dot{B}'-48 r^4 B^3 f F^4 \dot{B}'^2+8 r^3 B^4 f F^3 \dot{B} \dot{F}'-12 r^4 B^4 F^3 f' \dot{B} \dot{F}'-24 r^4 B^3 f F^3 B' \dot{B} \dot{F}'
       \nonumber
  \en
  \be
       +8 r^4 B^4 f F^2 F' \dot{B} \dot{F}'-40 r^3 B^5 f F^2 \dot{F} \dot{F}'+60 r^4 B^5 F^2 f' \dot{F} \dot{F}'-40 r^4 B^4 f F^2 B' \dot{F} \dot{F}'
       \nonumber
  \en
  \be
       -72 r^4 B^5 f F F' \dot{F} \dot{F}'+32 r^4 B^4 f F^3 \dot{B}' \dot{F}'+16 r^4 B^5 f F^2 \dot{F}'^2-24 r^4 B^3 f F^4 \dot{B} \dot{B}''
       \nonumber
  \en
  \be
       -8 r^4 B^4 f F^3 \dot{F} \dot{B}''-8 r^4 B^4 f F^3 \dot{B} \dot{F}''+40 r^4 B^5 f F^2 \dot{F} \dot{F}''+16 r^3 B^3 f F^4 B' \ddot{B}-24 r^4 B^3 F^4 f' B' \ddot{B}
       \nonumber
  \en
  \be
       +48 r^4 B^2 f F^4 B'^2 \ddot{B}-16 r^4 B^3 f F^3 B' F' \ddot{B}-16 r^4 B^3 f F^4 B'' \ddot{B}-16 r^3 B^5 f F^2 F' \ddot{F}+24 r^4 B^5 F^2 f' F' \ddot{F}
       \nonumber
  \en
  \be
       -16 r^4 B^4 f F^2 B' F' \ddot{F}-16 r^4 B^5 f F F'^2 \ddot{F}+16 r^4 B^5 f F^2 F'' \ddot{F}-16 r^3 B^4 f F^4 \ddot{B}'+24 r^4 B^4 F^4 f' \ddot{B}'-48 r^4 B^3 f F^4 B' \ddot{B}'
       \nonumber
  \en
  \be
       +16 r^4 B^4 f F^3 F' \ddot{B}'+16 r^3 B^5 f F^3 \ddot{F}'-24 r^4 B^5 F^3 f' \ddot{F}'+16 r^4 B^4 f F^3 B' \ddot{F}'+16 r^4 B^5 f F^2 F' \ddot{F}'+16 r^4 B^4 f F^4 \ddot{B}''
       \nonumber
  \en
  \be
     \left.  -16 r^4 B^5 f F^3 \ddot{F}''\right),
    \en

  \be
    8\pi p_\theta=8\pi p_\phi=\frac{\alpha}{24 r^4 B^6 f F^5}\left(-16 B^4 f F^5+16 B^4 f^3 F^5-32 r B^4 f^2 F^5 f'+16 r^2 B^4 f F^5 f'^2+16 r^2 B^4 f^2 F^5 f''\right.
    \nonumber
  \en
  \be
    -16 r^3 B^4 f F^5 f' f''+4 r^4 B^4 f F^5 f''^2-16 r^3 B^4 f^2 F^5 f'''-8 r^4 B^4 f F^5 f' f'''-8 r^4 B^4 f^2 F^5 f''''+16 r B^3 f^3 F^5 B'
    \nonumber
  \en
  \be
    -40 r^2 B^3 f^2 F^5 f' B'+40 r^3 B^3 f F^5 f'^2 B'+48 r^3 B^3 f^2 F^5 f'' B'+16 r^4 B^3 f^2 F^5 f''' B'+24 r^2 B^2 f^3 F^5 B'^2
    \nonumber
  \en
  \be
    -140 r^3 B^2 f^2 F^5 f' B'^2-15 r^4 B^2 f F^5 f'^2 B'^2-30 r^4 B^2 f^2 F^5 f'' B'^2+100 r^3 B f^3 F^5 B'^3+90 r^4 B f^2 F^5 f' B'^3-75 r^4 f^3 F^5 B'^4
    \nonumber
  \en
  \be
    -16 r B^4 f^3 F^4 F'+40 r^2 B^4 f^2 F^4 f' F'-40 r^3 B^4 f F^4 f'^2 F'-48 r^3 B^4 f^2 F^4 f'' F'-16 r^4 B^4 f^2 F^4 f''' F'-16 r^2 B^3 f^3 F^4 B' F'
    \nonumber
  \en
  \be
    +104 r^3 B^3 f^2 F^4 f' B' F'-10 r^4 B^3 f F^4 f'^2 B' F'+12 r^4 B^3 f^2 F^4 f'' B' F'-68 r^3 B^2 f^3 F^4 B'^2 F'-10 r^4 B^2 f^2 F^4 f' B'^2 F'
    \nonumber
  \en
  \be
    +20 r^4 B f^3 F^4 B'^3 F'-8 r^2 B^4 f^3 F^3 F'^2+36 r^3 B^4 f^2 F^3 f' F'^2+25 r^4 B^4 f F^3 f'^2 F'^2+18 r^4 B^4 f^2 F^3 f'' F'^2
    \nonumber
  \en
  \be
    -20 r^3 B^3 f^3 F^3 B' F'^2-34 r^4 B^3 f^2 F^3 f' B' F'^2+14 r^4 B^2 f^3 F^3 B'^2 F'^2-12 r^3 B^4 f^3 F^2 F'^3-46 r^4 B^4 f^2 F^2 f' F'^3
    \nonumber
  \en
  \be
    +20 r^4 B^3 f^3 F^2 B' F'^3+21 r^4 B^4 f^3 F F'^4-16 r^2 B^3 f^3 F^5 B''+88 r^3 B^3 f^2 F^5 f' B''+20 r^4 B^3 f F^5 f'^2 B''+24 r^4 B^3 f^2 F^5 f'' B''
    \nonumber
  \en
  \be
    -116 r^3 B^2 f^3 F^5 B' B''-130 r^4 B^2 f^2 F^5 f' B' B''+140 r^4 B f^3 F^5 B'^2 B''+44 r^3 B^3 f^3 F^4 F' B''+22 r^4 B^3 f^2 F^4 f' F' B''
    \nonumber
  \en
  \be
    -36 r^4 B^2 f^3 F^4 B' F' B''-8 r^4 B^3 f^3 F^3 F'^2 B''-28 r^4 B^2 f^3 F^5 B''^2+16 r^2 B^4 f^3 F^4 F''-88 r^3 B^4 f^2 F^4 f' F''
    \nonumber
  \en
  \be
    -20 r^4 B^4 f F^4 f'^2 F''-24 r^4 B^4 f^2 F^4 f'' F''+44 r^3 B^3 f^3 F^4 B' F''+22 r^4 B^3 f^2 F^4 f' B' F''-8 r^4 B^2 f^3 F^4 B'^2 F''
    \nonumber
  \en
  \be
    +28 r^3 B^4 f^3 F^3 F' F''+86 r^4 B^4 f^2 F^3 f' F' F''-36 r^4 B^3 f^3 F^3 B' F' F''-52 r^4 B^4 f^3 F^2 F'^2 F''+8 r^4 B^3 f^3 F^4 B'' F''
    \nonumber
  \en
  \be
    +20 r^4 B^4 f^3 F^3 F''^2+24 r^3 B^3 f^3 F^5 B'''+36 r^4 B^3 f^2 F^5 f' B'''-44 r^4 B^2 f^3 F^5 B' B'''+12 r^4 B^3 f^3 F^4 F' B'''
    \nonumber
  \en
  \be
    -24 r^3 B^4 f^3 F^4 F'''-36 r^4 B^4 f^2 F^4 f' F'''+12 r^4 B^3 f^3 F^4 B' F'''+20 r^4 B^4 f^3 F^3 F' F'''+8 r^4 B^3 f^3 F^5 B''''-8 r^4 B^4 f^3 F^4 F''''
    \nonumber
  \en
  \be
    +12 r^3 B^2 f F^4 B' \dot{B}^2-18 r^4 B^2 F^4 f' B' \dot{B}^2+60 r^4 B f F^4 B'^2 \dot{B}^2-12 r^4 B^2 f F^3 B' F' \dot{B}^2-12 r^4 B^2 f F^4 B'' \dot{B}^2
    \nonumber
  \en
  \be
    +4 r^3 B^3 f F^3 B' \dot{B} \dot{F}-6 r^4 B^3 F^3 f' B' \dot{B} \dot{F}+12 r^4 B^2 f F^3 B'^2 \dot{B} \dot{F}+4 r^3 B^4 f F^2 F' \dot{B} \dot{F}-6 r^4 B^4 F^2 f' F' \dot{B} \dot{F}
    \nonumber
  \en
  \be
    -16 r^4 B^3 f F^2 B' F' \dot{B} \dot{F}+4 r^4 B^4 f F F'^2 \dot{B} \dot{F}-4 r^4 B^3 f F^3 B'' \dot{B} \dot{F}-4 r^4 B^4 f F^2 F'' \dot{B} \dot{F}-20 r^3 B^5 f F F' \dot{F}^2
    \nonumber
  \en
  \be
    +30 r^4 B^5 F f' F' \dot{F}^2-20 r^4 B^4 f F B' F' \dot{F}^2-28 r^4 B^5 f F'^2 \dot{F}^2+20 r^4 B^5 f F F'' \dot{F}^2-12 r^3 B^3 f F^4 \dot{B} \dot{B}'+18 r^4 B^3 F^4 f' \dot{B} \dot{B}'
    \nonumber
  \en
  \be
    -84 r^4 B^2 f F^4 B' \dot{B} \dot{B}'+12 r^4 B^3 f F^3 F' \dot{B} \dot{B}'-4 r^3 B^4 f F^3 \dot{F} \dot{B}'+6 r^4 B^4 F^3 f' \dot{F} \dot{B}'-12 r^4 B^3 f F^3 B' \dot{F} \dot{B}'
    \nonumber
  \en
  \be
    +20 r^4 B^4 f F^2 F' \dot{F} \dot{B}'+24 r^4 B^3 f F^4 \dot{B}'^2-4 r^3 B^4 f F^3 \dot{B} \dot{F}'+6 r^4 B^4 F^3 f' \dot{B} \dot{F}'+12 r^4 B^3 f F^3 B' \dot{B} \dot{F}'
    \nonumber
  \en
  \be
    -4 r^4 B^4 f F^2 F' \dot{B} \dot{F}'+20 r^3 B^5 f F^2 \dot{F} \dot{F}'-30 r^4 B^5 F^2 f' \dot{F} \dot{F}'+20 r^4 B^4 f F^2 B' \dot{F} \dot{F}'+36 r^4 B^5 f F F' \dot{F} \dot{F}'
    \nonumber
  \en
  \be
    -16 r^4 B^4 f F^3 \dot{B}' \dot{F}'-8 r^4 B^5 f F^2 \dot{F}'^2+12 r^4 B^3 f F^4 \dot{B} \dot{B}''+4 r^4 B^4 f F^3 \dot{F} \dot{B}''+4 r^4 B^4 f F^3 \dot{B} \dot{F}''
    \nonumber
  \en
  \be
    -20 r^4 B^5 f F^2 \dot{F} \dot{F}''-8 r^3 B^3 f F^4 B' \ddot{B}+12 r^4 B^3 F^4 f' B' \ddot{B}-24 r^4 B^2 f F^4 B'^2 \ddot{B}+8 r^4 B^3 f F^3 B' F' \ddot{B}+8 r^4 B^3 f F^4 B'' \ddot{B}
    \nonumber
  \en
  \be
    +8 r^3 B^5 f F^2 F' \ddot{F}-12 r^4 B^5 F^2 f' F' \ddot{F}+8 r^4 B^4 f F^2 B' F' \ddot{F}+8 r^4 B^5 f F F'^2 \ddot{F}-8 r^4 B^5 f F^2 F'' \ddot{F}+8 r^3 B^4 f F^4 \ddot{B}'
    \nonumber
  \en
  \be
    -12 r^4 B^4 F^4 f' \ddot{B}'+24 r^4 B^3 f F^4 B' \ddot{B}'-8 r^4 B^4 f F^3 F' \ddot{B}'-8 r^3 B^5 f F^3 \ddot{F}'+12 r^4 B^5 F^3 f' \ddot{F}'-8 r^4 B^4 f F^3 B' \ddot{F}'
    \nonumber
  \en
  \be
    \left.-8 r^4 B^5 f F^2 F' \ddot{F}'-8 r^4 B^4 f F^4 \ddot{B}''+8 r^4 B^5 f F^3 \ddot{F}''\right),
    \en

    \be
    0=T^1_0=\frac{\alpha}{6 r^2 B^6 F^4}\left(-8 B^2 f^2 F^4 B' \dot{B}+16 r B^2 f F^4 f' B' \dot{B}-3 r^2 B^2 F^4 f'^2 B' \dot{B}+6 r^2 B^2 f F^4 f'' B' \dot{B}
     \right.\nonumber
  \en
  \be
    -30 r B f^2 F^4 B'^2 \dot{B}-15 r^2 B f F^4 f' B'^2 \dot{B}+30 r^2 f^2 F^4 B'^3 \dot{B}+14 r B^2 f^2 F^3 B' F' \dot{B}-r^2 B^2 f F^3 f' B' F' \dot{B}
     \nonumber
  \en
  \be
    -6 r^2 B^2 f^2 F^2 B' F'^2 \dot{B}+8 r B^2 f^2 F^4 B'' \dot{B}+8 r^2 B^2 f F^4 f' B'' \dot{B}-30 r^2 B f^2 F^4 B' B'' \dot{B}+2 r^2 B^2 f^2 F^3 F' B'' \dot{B}
     \nonumber
  \en
  \be
    +4 r^2 B^2 f^2 F^3 B' F'' \dot{B}+4 r^2 B^2 f^2 F^4 B''' \dot{B}+8 B^4 f^2 F^2 F' \dot{F}-16 r B^4 f F^2 f' F' \dot{F}+3 r^2 B^4 F^2 f'^2 F' \dot{F}
     \nonumber
  \en
  \be
    -6 r^2 B^4 f F^2 f'' F' \dot{F}+14 r B^3 f^2 F^2 B' F' \dot{F}-r^2 B^3 f F^2 f' B' F' \dot{F}-2 r^2 B^2 f^2 F^2 B'^2 F' \dot{F}+2 r B^4 f^2 F F'^2 \dot{F}
     \nonumber
  \en
  \be
    +17 r^2 B^4 f F f' F'^2 \dot{F}-8 r^2 B^3 f^2 F B' F'^2 \dot{F}-14 r^2 B^4 f^2 F'^3 \dot{F}+4 r^2 B^3 f^2 F^2 F' B'' \dot{F}-8 r B^4 f^2 F^2 F'' \dot{F}
     \nonumber
  \en
  \be
    -8 r^2 B^4 f F^2 f' F'' \dot{F}+2 r^2 B^3 f^2 F^2 B' F'' \dot{F}+18 r^2 B^4 f^2 F F' F'' \dot{F}-4 r^2 B^4 f^2 F^2 F''' \dot{F}+8 B^3 f^2 F^4 \dot{B}'
     \nonumber
  \en
  \be
    -16 r B^3 f F^4 f' \dot{B}'+3 r^2 B^3 F^4 f'^2 \dot{B}'-6 r^2 B^3 f F^4 f'' \dot{B}'+30 r B^2 f^2 F^4 B' \dot{B}'+15 r^2 B^2 f F^4 f' B' \dot{B}'
     \nonumber
  \en
  \be
    -30 r^2 B f^2 F^4 B'^2 \dot{B}'-14 r B^3 f^2 F^3 F' \dot{B}'+r^2 B^3 f F^3 f' F' \dot{B}'+6 r^2 B^3 f^2 F^2 F'^2 \dot{B}'+16 r^2 B^2 f^2 F^4 B'' \dot{B}'
     \nonumber
  \en
  \be
    -4 r^2 B^3 f^2 F^3 F'' \dot{B}'-8 B^4 f^2 F^3 \dot{F}'+16 r B^4 f F^3 f' \dot{F}'-3 r^2 B^4 F^3 f'^2 \dot{F}'+6 r^2 B^4 f F^3 f'' \dot{F}'-14 r B^3 f^2 F^3 B' \dot{F}'
     \nonumber
  \en
  \be
    +r^2 B^3 f F^3 f' B' \dot{F}'+2 r^2 B^2 f^2 F^3 B'^2 \dot{F}'-2 r B^4 f^2 F^2 F' \dot{F}'-17 r^2 B^4 f F^2 f' F' \dot{F}'+8 r^2 B^3 f^2 F^2 B' F' \dot{F}'
     \nonumber
  \en
  \be
    +14 r^2 B^4 f^2 F F'^2 \dot{F}'-4 r^2 B^3 f^2 F^3 B'' \dot{F}'-8 r^2 B^4 f^2 F^2 F'' \dot{F}'-8 r B^3 f^2 F^4 \dot{B}''-8 r^2 B^3 f F^4 f' \dot{B}''
     \nonumber
  \en
  \be
    +14 r^2 B^2 f^2 F^4 B' \dot{B}''-2 r^2 B^3 f^2 F^3 F' \dot{B}''+8 r B^4 f^2 F^3 \dot{F}''+8 r^2 B^4 f F^3 f' \dot{F}''-2 r^2 B^3 f^2 F^3 B' \dot{F}''
     \nonumber
  \en
  \be
    \left.-10 r^2 B^4 f^2 F^2 F' \dot{F}''-4 r^2 B^3 f^2 F^4 \dot{B}'''+4 r^2 B^4 f^2 F^3 \dot{F}'''\right).
    \en


\begin{thebibliography}{99}

   \bibitem{wey}
    H. Weyl, $Space~ Time~ Matter$, Dover Publications, New York, 1950.

 \bibitem{olz}
  E.~Scholz,
  arXiv:1111.3220 [math.HO].

\bibitem{man1}
  P.~D.~Mannheim,
  Found.\ Phys.\  {\bf 42}, 388 (2012)
  [arXiv:1101.2186 [hep-th]].

\bibitem{man2}
  P.~D.~Mannheim and D.~Kazanas,
  Astrophys.\ J.\  {\bf 342}, 635 (1989).

\bibitem{man3}
  P.~D.~Mannheim,
  Phys.\ Rev.\ D {\bf 85}, 124008 (2012)
  doi:10.1103/PhysRevD.85.124008
  [arXiv:1109.4119 [gr-qc]].
\bibitem{yang}
  R.~Yang, B.~Chen, H.~Zhao, J.~Li and Y.~Liu,
  Phys.\ Lett.\ B {\bf 727}, 43 (2013)
  doi:10.1016/j.physletb.2013.10.035
  [arXiv:1311.2800 [gr-qc]].

   \bibitem{gru}
  D.~Grumiller, M.~Irakleidou, I.~Lovrekovic and R.~McNees,
  Phys.\ Rev.\ Lett.\  {\bf 112}, 111102 (2014)
  doi:10.1103/PhysRevLett.112.111102
  [arXiv:1310.0819 [hep-th]].

  \bibitem{des}
  S.~Deser and B.~Tekin,
  Phys.\ Rev.\ D {\bf 67}, 084009 (2003)
  doi:10.1103/PhysRevD.67.084009
  [hep-th/0212292].
  \bibitem{hoo}
  G.~'t Hooft,
  Found.\ Phys.\  {\bf 41}, 1829 (2011)
  doi:10.1007/s10701-011-9586-8
  [arXiv:1104.4543 [gr-qc]].

\bibitem{mal}
  J.~Maldacena,
  arXiv:1105.5632 [hep-th].
\bibitem{liu}
  H.~S.~Liu and H.~Lu,
  JHEP {\bf 1302} (2013) 139
  doi:10.1007/JHEP02(2013)139
  [arXiv:1212.6264 [hep-th]]; H.~S.~Liu, H.~L¨¹, C.~N.~Pope and J.~F.~V¨¢zquez-Poritz,
  Class.\ Quant.\ Grav.\  {\bf 30}, 165015 (2013)
  doi:10.1088/0264-9381/30/16/165015
  [arXiv:1303.5781 [hep-th]].

  \bibitem{zhao}
   K.~Meng and L.~Zhao,
  arXiv:1601.07634 [gr-qc].

 \bibitem{said}
  J.~L.~Said, J.~Sultana and K.~Z.~Adami,
  Phys.\ Rev.\ D {\bf 85}, 104054 (2012)
  doi:10.1103/PhysRevD.85.104054
  [arXiv:1201.0860 [gr-qc]];  J.~L.~Said, J.~Sultana and K.~Z.~Adami,
  Phys.\ Rev.\ D {\bf 86}, 104009 (2012)
  doi:10.1103/PhysRevD.86.104009
  [arXiv:1207.2108 [gr-qc]].
  \bibitem{nar}
  J. Narlikar  and A. Kembhavi,  Lett. Nuovo Cim. 19 517(1977).
  \bibitem{bambi1}
  C.~Bambi, L.~Modesto and L.~Rachwal,
  arXiv:1611.00865 [gr-qc].
  \bibitem{bambi2}
  C.~Bambi, L.~Modesto, S.~Porey and L.~Rachwal,
  arXiv:1611.05582 [gr-qc].
  \bibitem{modesto}
  L.~Modesto and L.~Rachwal,
  arXiv:1605.04173 [hep-th].
  \bibitem{mcv}
  G. C. McVittie, Mon. Not. R. Astron. Soc. 93, 325 (1933); Mon. Not. R. Astron. Soc. 92, 500 (1932).

  \bibitem{gao}
    C.~J.~Gao and S.~N.~Zhang,
  Phys.\ Lett.\ B {\bf 595}, 28 (2004)
  doi:10.1016/j.physletb.2004.05.076
  [gr-qc/0407045]; C.~J.~Gao,
  Class.\ Quant.\ Grav.\  {\bf 21}, 4805 (2004)
  doi:10.1088/0264-9381/21/21/004
  [gr-qc/0411033]; C.~J.~Gao and S.~N.~Zhang,
  Gen.\ Rel.\ Grav.\  {\bf 38}, 23 (2006)
  doi:10.1007/s10714-005-0207-8
  [gr-qc/0411040].
\bibitem{mil}
  E. A. Milne, ¡°World-Structure and the Expansion of the
Universe¡±, Z.Astrophysik 6: 1-35 (1933).

  \bibitem{nie}
  J.~T.~Nielsen, A.~Guffanti and S.~Sarkar,
  Sci.\ Rep.\  {\bf 6}, 35596 (2016)
  doi:10.1038/srep35596
  [arXiv:1506.01354 [astro-ph.CO]].

 \bibitem{bir}
  R.~J.~Riegert,
  Phys.\ Rev.\ Lett.\  {\bf 53}, 315 (1984).
  doi:10.1103/PhysRevLett.53.315
  \bibitem{self1}
   H.~Zhang and X.~Z.~Li,
  Phys.\ Lett.\ B {\bf 737}, 395 (2014)
  doi:10.1016/j.physletb.2014.09.010
  [arXiv:1406.1553 [gr-qc]]; H.~Zhang, Y.~Hu and X.~Z.~Li,
  Phys.\ Rev.\ D {\bf 90}, no. 2, 024062 (2014)
  doi:10.1103/PhysRevD.90.024062
  [arXiv:1406.0577 [gr-qc]];  H.~Zhang, S.~A.~Hayward, X.~H.~Zhai and X.~Z.~Li,
  Phys.\ Rev.\ D {\bf 89}, no. 6, 064052 (2014)
  doi:10.1103/PhysRevD.89.064052
  [arXiv:1304.3647 [gr-qc]]; H.~Zhang,
  The Universe {\bf 3} (2015) no.1,  30; D.~He and Q.~y.~Cai,
  arXiv:1609.05825 [hep-th]; H.~W.~Tan, J.~B.~Yang, T.~M.~He and J.~Y.~Zhang,
  arXiv:1609.04181 [gr-qc].


  \bibitem{self2}
 H.~Zhang and X.~Z.~Li,
  Phys.\ Lett.\ B {\bf 700}, 97 (2011).
  doi:10.1016/j.physletb.2011.04.058;  H.~Zhang and H.~Noh,
  Phys.\ Lett.\ B {\bf 671}, 428 (2009)
  doi:10.1016/j.physletb.2008.12.057
  [arXiv:0904.0065 [gr-qc]];H.~Zhang and H.~Noh,
  Phys.\ Lett.\ B {\bf 670}, 271 (2009)
  doi:10.1016/j.physletb.2008.11.015
  [arXiv:0904.0063 [gr-qc]];H.~s.~Zhang, H.~Noh and Z.~H.~Zhu,
  Phys.\ Lett.\ B {\bf 663}, 291 (2008)
  doi:10.1016/j.physletb.2008.04.022
  [arXiv:0804.2931 [gr-qc]].

























  \bibitem{mel}
   M. A. Melvin, Phys. Lett. 8 (1964) 65.

\end{thebibliography}
\end{document}